\documentclass[twoside]{article}
\usepackage{epsfig.sty}
\newcommand{\beq}{\begin{eqnarray}}
\newcommand{\eeq}{\end{eqnarray}}
\begin{document}
\title{Gravitational Waves Created During the EWPT}
\author{Leonard S. Kisslinger \\
Department of Physics, Carnegie Mellon University, Pittsburgh, PA 15213}
\date{}
\maketitle
\begin{abstract}We study gravitational waves generated by bubble expansion
  created during the Cosmological Electroweak Phase Transition (EWPT). 
  The energy radiated via gravitational waves is produced by the stress-energy
  tensor created by the magnetic field produced by bubble collisions during
  the EWPT, which occured about $10{-11}$ seconds after the Big Bang.

\end{abstract}
  
\noindent
PACS Indices:98.70.~Vc, 98.80.-k, 98-62.~En, 98.80.~Cq
\vspace{1mm}

\noindent
Keywords:Gravitational Waves,energy radiated, Electroweak Phase Transition
\section{Introduction}

The present work is an extension of the estimate of gravitational waves
created during the Cosmological Quantum Chromodynamic Phase transition
(QCDPT)\cite{lsk19} to an estimate of gravitational waves created during the
Cosmological Electroweak Phase Transition (EWPT). It is also closely related
to a recent publication of gravitational radiation produced by pulsar
creation\cite{lsz18}.

The QCDPT and EWPT are first order cosmological phase transitions so they
have latent heat: the  quark condensate for the QCDPT and the Higgs mass for
the EWPT.

For the QCDPT $q(x),\bar{q}(x)$ are the quark, antiquark fields. $| >$ is the
vacuum state.
\beq
< |\bar{q}(x) q(x)| > &=& {\rm \;quark\;condensate} \nonumber \\
< |\bar{q}(x) q(x)| >&=& 0 {\rm \;in\;quark\;gluon\;plasma\;phase}
\nonumber \\
           &\simeq& -(.23\;GeV)^3 {\rm \;in\;hadron\;phase} \nonumber \; .
\eeq
Therefore for the QCDPT the latent heat $\simeq -(.23\;GeV)^3$.

For the EWPT, with $\phi_H$ the Higgs field, with $T_c\simeq$ 125 GeV the
critical temperature,
\beq
< |\phi_H| > &=& 0 {\rm \;\;\;for\; T_c} \geq 125 {\rm \;GeV} \nonumber \\
< |\phi_H| > &=& 125 GeV {\rm \;\;\;for\; T_c } \leq 125 {\rm \;GeV}
\nonumber \; .
\eeq
Therefore for the EWPT the latent heat $\simeq$ 125 GeV=$M_H$, the Higgs mass.

There have been several articles published on gravitational radiation produced
by Cosmological Phase Transitions: the Electroweak Phase Transition (EWPT) at
about $10{-11}$ seconds after the Big Bang and the Quantum Chromodynamics Phase
Transition (QCDPT) at about $10^{-4}$ seconds.

One of the first studies, which is a basis for the present work, was
``Gravitational radiation from first-order phase transitions'' by Kamionkowski,
Kosowsky and Turner\cite{kkt94}. More recent studies were gravitational
radiation from primoidal turbulence\cite{kcgmr}, gravitationl radiation from
cosmological phase ransition magnetic fields\cite{kks10}, and polarization of
such gravitational radiation\cite{kk15}.

The present work also makes use of the stress-energy tensor produced by
the magnetic wall during the EWPT\cite{lsk03}. In order to estimate the
energy radiated by gravitational waves during the EWPT one needs the
nucleon mass $M_n$ in units of $fm^{-1}$, $B_{W}$, the magnitude of the
magnetic field at the bubble wall during the EWPT, and $t_{EWPT}=10^{-11}$
seconds after the Big Bang, the time of the EWPT. The values of these parameters
used in the present work are taken from estimates in Ref\cite{lsk03}.

The only experimental detection of gravitational waves was the observation
of  gravitational waves from binary black hole mergers Ref\cite{ligo}. Gravity
waves from black hole mergers had been predicted\cite{bak06,cam06} and were
in agreement within experimental and theoretical errors with Ref\cite{ligo}.
\newpage
\section{Electroweak Phase Transition (EWPT)}
The Standard EW Model has fields with quanta:

Fermions (quantum spin 1/2 particles) are $e^-,\nu_e$, the
$\mu$ and $\tau$ leptons, the quarks $q_u,q_d$ and  two other quark generations.
The EW gauge bosons (quantum spin 1) are $W^+,W^-,Z^o$ and photon $\gamma$

Since the EWPT is a first-order phase transition there is latent heat. The
latent heat for the EWPT is the Higgs boson (quantum spin 0) mass, $M_H=$.
125 GeV. At the LHC\cite{lhc} is was found that $M_H \simeq$ 125 GeV. Also
during the EWPT the quarks caused Baryogenesis, with more particles than
antiparticles, due to CP (Charge Congugation and Parity) violation

The EW diagrams are shown in the figure below.
\begin{figure}[ht]
\begin{center}
\epsfig{file=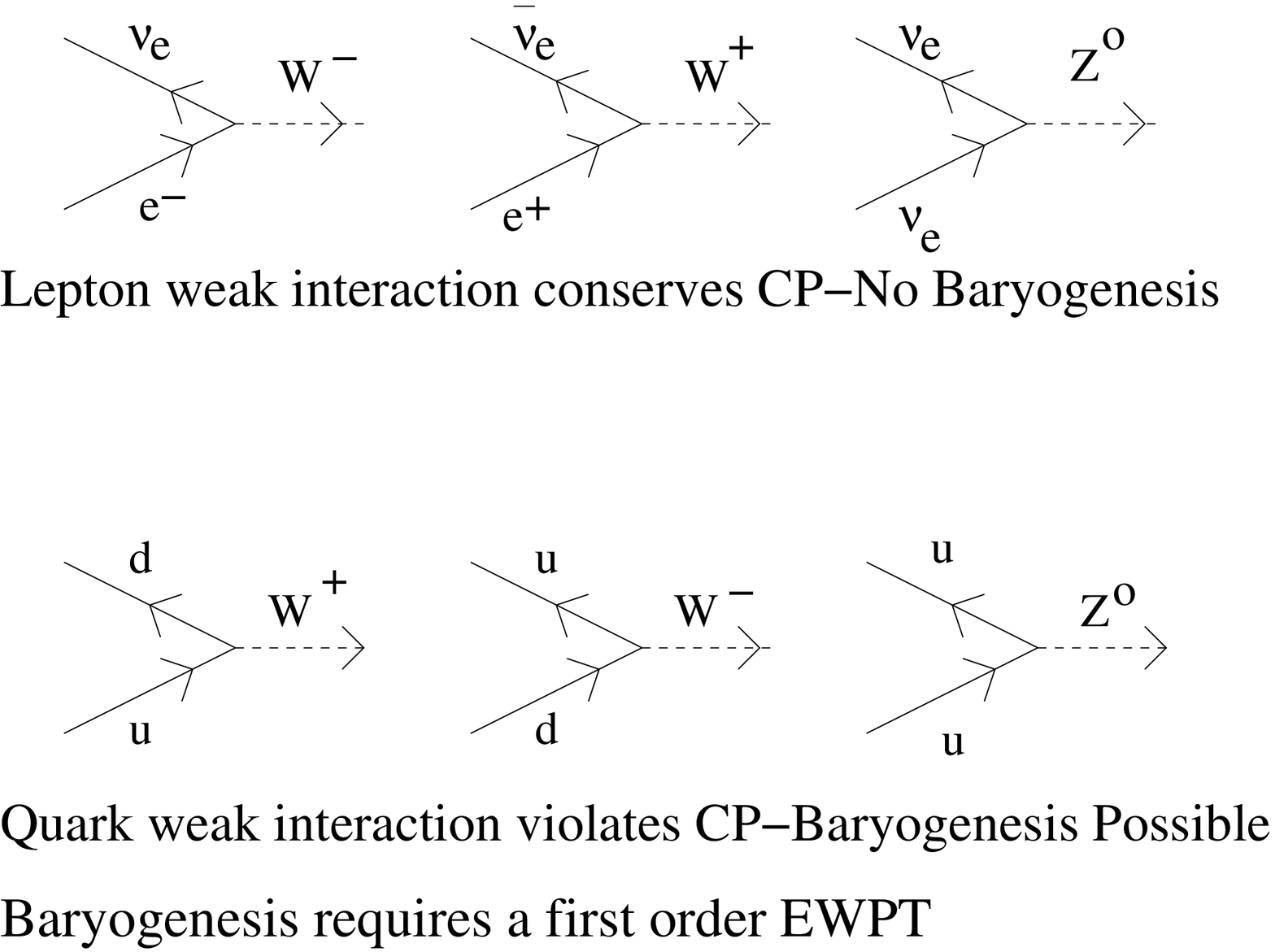,height=8cm,width=12cm}
\label{Fig.1}
\end{center}
\end{figure}

During the EWPT bubbles form and magnetic fields were created via
bubble collisions, as shown in the figure below.
\begin{figure}[ht]
\begin{center}
\epsfig{file=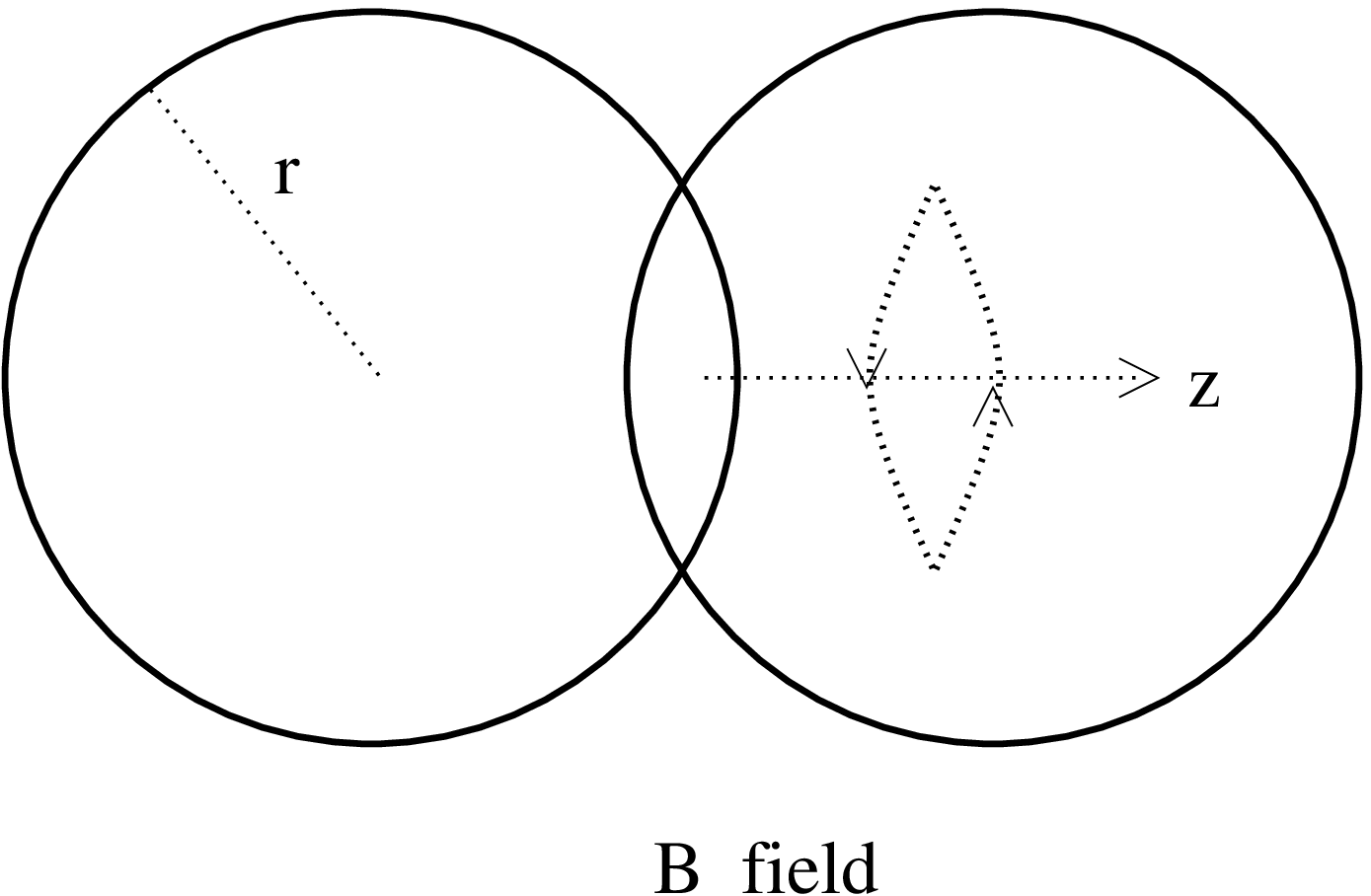,height=6cm,width=10cm}
\label{Fig.2}
\end{center}
\end{figure}

From Ref\cite{kks10}, $B_W \simeq 4.3 \times 10^{24} {\rm Gauss}$. Therefore
$B_W^{EPPT} \simeq 4.3 \times 10^7 \times B_W^{QCDPT}$.

\newpage

\section{Gravitational Radiation From Magnetic Fields
  Generated by the EWPT} 
The energy radiated by gravitational waves with frequency interval $d\omega$
and solid angle $d\Omega$ is\cite{kkt94}

\beq
\label{dEdomegaOmega}
\frac{dE}{d\omega d\Omega}&=& 2G\omega^2\Lambda_{ij,lm}(\hat{k})
T^*_{ij}(k,\omega)T_{lm}(k,\omega) \; ,
\eeq

\beq
\label{dEdomegaOmega2}
\Lambda_{ij,lm}(\hat{k})&=& \delta_{il}\delta_{jm}-\delta_{ij}\delta_{lm}/2
+\delta_{ij}\hat{k}_l\hat{k}_m/2 +\delta_{lm}\hat{k}_i\hat{k}_j/2
-2\delta_{il}\hat{k}_j \hat{k}_m +\hat{k}_i \hat{k}_j \hat{k}_l \hat{k}_m/2
 \; .
\eeq

From Eq(20) in Ref\cite{lsk03}, which makes use of Eq(\ref{dEdomegaOmega}),
with $k_3\simeq k/\sqrt{3}$, 
\beq
\label{Tij}
T_{ij}(k,\omega)&=&\delta_{i3}\delta_{j3}\frac{2 \pi^3 \sqrt{\pi}}{M_n}
B_W^2 e^{3k_3/8M-n^2}\delta(t-t_{EWPT}) \; ,
\eeq
where $t_{EWPT}$ is the time of the EWPT, with
\beq
\label{tQ-BW-MN}
t_{EWPT}&\simeq& 10^{-11} {\rm \;\;seconds} \\
B_W&\simeq& 4.3 \times 10^{24} {\rm \;\;Gauss} \nonumber \\
M_N^{-1}&\simeq& 0.2 {\rm \;\;fm} \nonumber \; .
\eeq

From Eqs(\ref{dEdomegaOmega},\ref{dEdomegaOmega2},\ref{Tij}) 
the energy radiated by gravitational waves with frequency interval $d \omega$,
eliminating the solid angle as in Ref\cite{kkt94}, using $k_3 \simeq k/\sqrt{3}$
and \cite{kks10}
\beq
\label{dEdomegaOmega3}
\frac{dE}{d\omega}&=& \frac{8G\omega^2 \pi^7 B_W^4}{M_N^2}
e^{-3k_3/4M-n^2}[\frac{1}{2}-\frac{k_3^2}{k^2}+\frac{k_3^4}{2 k^4}] \; .
\eeq
since $k_3\simeq k/\sqrt{3}$ or $k_3/k \simeq 1/\sqrt{3}$, from
Eq(\ref{dEdomegaOmega3}) one finds
\beq
\label{dEdomegaOmega4}
\frac{dE}{d\omega}&\simeq& \frac{8G\omega^2 \pi^7 B_W^4}{M_N^2}
e^{-\frac{k^2}{18 M_N^2}} \; .
\eeq  
From Refs(\cite{kks10},\cite{kkt94},\cite{bvs06})
\beq
\label{k}
k &\simeq& 2\pi/\lambda_o \simeq 2\pi/(10^{13} fm)   \; .
\eeq

Therefore, using $M_N^{-1} \simeq 0.2$ fm
\beq
\label{exp}
e^{-k^2/4M_N^2}&\simeq& e^{0} =1.0 /; ,
\eeq
from Eq(\ref{dEdomegaOmega4})

\beq
\label{dEdomegaOmega5}
\frac{dE}{d\omega}&=& \frac{8G\omega^2 \pi^7 B_W^4}{M_N^2} \; .
\eeq

Using (with \rm{s}=second)
\beq
\label{GBW}
G(\rm \;Gauss)^4 &=& \frac{8.093 10^{-47}}{\rm \;s^2\;cm^2} \\
{\rm cm}&=& 10^{13} {\rm \;fm} \; ,
\eeq
and Eq(\ref{tQ-BW-MN}) one obtains our final equation for the energy radiated
by gravitational waves during the QCDPT
\beq
\label{dEdomega}
\frac{dE}{d\omega}&=& 3.36 10^{6} \frac{\omega^2}{{\rm\;s^2}} \; .
\eeq
\newpage

A review of Gravitational Wave Physics\cite{k02} discusses the generation of
gravitational waves, the first detection of Gravitational waves by the LIGO
detector\cite{LIGO}, and many other aspects of Gravitational Wave Physics.
An aspect of Ref\cite{k02} important for the present work is that the units
for Gravitational radiation are explained.

These are used for $\frac{dE}{d\omega}$ shown in Figure 1, with $s=$ second. 

\vspace{1.5cm}
\begin{figure}[ht]
\begin{center}
\epsfig{file=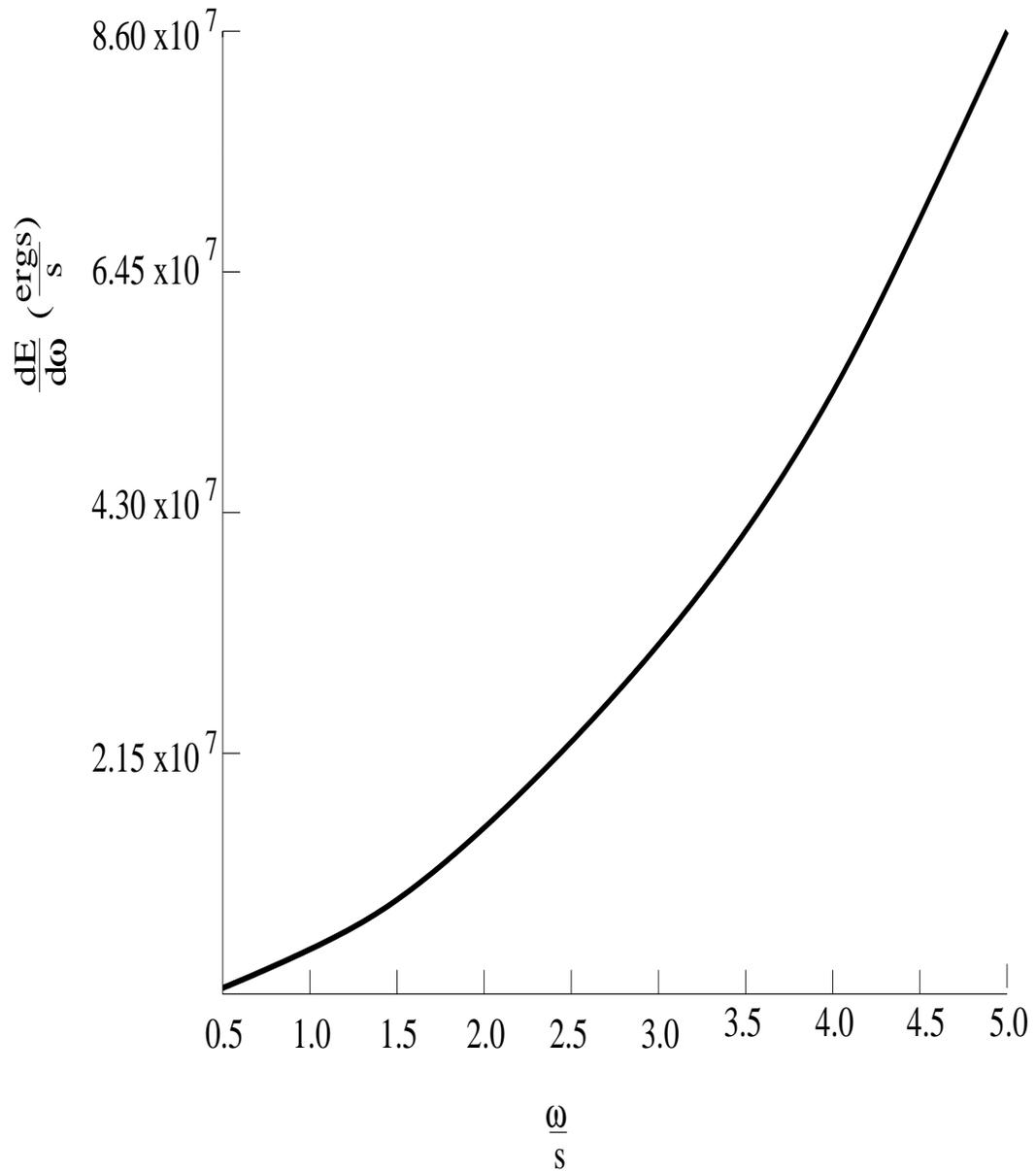,height=16cm,width=14cm}
\end{center}
\caption{Gravitational radiation energy produced during the QCDPT as a
 function of frequency $\omega$.}
{\label{Figure 1}}
\end{figure}
\newpage

\section{Conclusions}

We have estimated the gravitational radiation energy produced by gravitational
waves during the EWPT, which occured at a time $t_{EWPT} \simeq 10^{-11}$
seconds after the Big Bang. This estimate is based on methods found in
Ref\cite{kkt94}, but the estimate is of gravitational radiation energy
produced during a specific Cosmological Phase Transition, the Electroweak
Phase Transition, rather than a more general study of
Gravitational radiation from first-order phase transitions\cite{kkt94}.

From Figure 1 the Gravitational radiation energy as a function of $\omega$
for the most likely values of $\omega/s$ are approximately 2$\times 10^7$ to
8$\times 10^7$ ergs/s. This is approximately 4.3$\times 10^7$ larger than
the Gravitational radiation energy produced during the QCDPT\cite{lsk19}.

Note that gravitational waves from binary black hole mergers were detected
in 2016\cite{ligo}. From the results shown in Figure 1, with  Gravitational
radiation energy $\simeq 8\times 10^7$ ergs/s the current gravitational wave
detectors, such as LIGO\cite{LIGO}, can detect gravitational waves produced
during the EWPT. We hope that such LIGO experiments are carried out soon. 

\vspace{1cm}
\Large{{\bf Acknowledgements}}\\
\normalsize

 The author Leonard S. Kisslinger was a visitor at Los Alamos National 
Laboratory, Group P25.


\begin{thebibliography}{99}
\bibitem{lsk19}Leonard S. Kisslinger, Int. J. Mod. Phys. E {\bf D 19}, 00129
  (2019)
\bibitem{lsz18}Leonard S. Kisslinger, Bijit Singha and Zhou Li-juan,
  arXiv:1808.05265[astro-ph] (2018)
\bibitem{kkt94}Mark Kamionkowski, Authur Kosowsky and Michael S. Turner, Phys.
  Rev. {\bf D 49}, 2837 (1994)
\bibitem{kcgmr}Tina Kahniashvili, Leonardo Campanelli, Grigol Gogoberidze,
  Yurii Maravin and Bharat Ratra, Phys. Rev. {\bf D 78}, 12300 (2008) 
\bibitem{kks10}Tina Kahniashvili, Leonard Kisslinger and Trevor Stevens,
Phys. Rev. {\bf D 81}, 023004 (2010)  
\bibitem{kk15}Leonard Kisslinger and Tina Kahniashvili, Phys. Rev. {\bf D 92},
  043006 (2015)
\bibitem{lsk03}Leonard S. Kisslinger, Phys. Rev. {\bf D 68}, 0431516 (2003)
\bibitem{ligo}B.D. Abott $et.$$al.$, Ligo ScientificCollaboration and Virgo
  Scientific Collaboration, Phys. Rev. Lett. {\bf 116}, 061102 (2016)
\bibitem{bak06} Baker $et.$$al.$, Rev. Lett. {\bf 96}, 111102 (2006)
\bibitem{cam06} Campane $et.$$al.$, Rev. Lett. {\bf 96}, 111101 (2006)
\bibitem{lhc}The ALICE Collaboration, K Aamodt1 et.al. J. of Instrumentation
 {\bf 3} (2008)
\bibitem{bvs06}D. Boyanovsky, H.I. deVega and P.J. Schwartz, Ann. Rev. Nucl.
  Part. Sci. {\bf 56}, 441 (2006)
\bibitem{k02}Kostas D. Kokkotas, Encyclopedia of Physical Science and
  Technology, 3rd Edition, Volume 7 Academic Press (2002)
\bibitem{LIGO} LIGO, http://ligo.caltech.edu

\end{thebibliography}
\end{document}